\documentclass[sigconf,screen]{acmart}
\AtBeginDocument{%
  }

\acmConference[Preprint]{}{Preprint}{2025}
\setcopyright{none} %remove copyright notice
\acmISBN{}
\acmDOI{}

\DeclareMathOperator{\mel2hz}{Mel2Hz}
\DeclareMathOperator{\hz2mel}{Hz2Mel}

\begin{document}

% \title{Explicit Bandwidth Extension Prior to Vocoder for Singing Voice Synthesis with Realistic Spectrogram}
\title{Enhancing Spectrogram Realism in Singing Voice Synthesis via Explicit Bandwidth Extension Prior to Vocoder}

\author{Runxuan Yang}
% \email{yangrx20@mails.tsinghua.edu.cn}
\affiliation{%
  \institution{Tsinghua University}
  \city{Beijing}
  % \state{Beijing}
  \country{China}}

\author{Kai Li}
\affiliation{%
  \institution{Tsinghua University}
  \city{Beijing}
  % \state{Beijing}
  \country{China}}

\author{Guo Chen}
\affiliation{%
  \institution{Tsinghua University}
  \city{Beijing}
  % \state{Beijing}
  \country{China}}

\author{Xiaolin Hu}
\authornote{Corresponding author.}
\affiliation{%
 \institution{Tsinghua University}
  \city{Beijing}
  % \state{Beijing}
  \country{China}}

% \renewcommand{\shortauthors}{Yang et al.}

\begin{abstract}
This paper addresses the challenge of enhancing the realism of vocoder-generated singing voice audio by mitigating the distinguishable disparities between synthetic and real-life recordings, particularly in high-frequency spectrogram components. Our proposed approach combines two innovations: an explicit linear spectrogram estimation step using denoising diffusion process with DiT-based neural network architecture optimized for time-frequency data, and a redesigned vocoder based on Vocos specialized in handling large linear spectrograms with increased frequency bins. This integrated method can produce audio with high-fidelity spectrograms that are challenging for both human listeners and machine classifiers to differentiate from authentic recordings. Objective and subjective evaluations demonstrate that our streamlined approach maintains high audio quality while achieving this realism. This work presents a substantial advancement in overcoming the limitations of current vocoding techniques, particularly in the context of adversarial attacks on fake spectrogram detection.

\end{abstract}

\iffalse
\begin{teaserfigure}
  \includegraphics[width=\textwidth]{sampleteaser}
  \caption{Seattle Mariners at Spring Training, 2010.}
  \Description{Enjoying the baseball game from the third-base
  seats. Ichiro Suzuki preparing to bat.}
  \label{fig:teaser}
\end{teaserfigure}

\received{20 February 2007}
\received[revised]{12 March 2009}
\received[accepted]{5 June 2009}
\fi

\maketitle

\section{Introduction}

Recent advancements in deep learning have significantly elevated the quality and expressiveness of singing voice synthesis, transforming how synthetic audio is generated and perceived. These pipelines typically comprises two primary stages. First, symbolic tokens like phonemes are input into an acoustic model, generating acoustic features like Mel spectrograms or other spectral representations. These features are then fed into a vocoder, which synthesizes the final audio waveform. While this approach has significantly improved perceived listening experience, noticeable gaps persist between vocoder-generated audio and real-life recordings, making synthesized audio easily distinguishable from real ones.

A key distinction between them lies in the spectrogram, especially in the higher frequencies. Authentic recordings reveal distinct energy bands in their spectrograms reflecting the nuanced characteristics of the human voice. In contrast, vocoder-generated audio often displays blurry and overly smoothed high-frequency components. This discrepancy stems from multiple factors. Mel spectrograms, a common input to vocoders, inherently offer reduced resolution at higher frequencies due to their logarithmic frequency scaling. Moreover, many neural vocoders impose a maximum frequency cap—typically 8 kHz or 12 kHz—forcing them to reconstruct both phase information and high-frequency components simultaneously. However, existing vocoders predominantly prioritize phase reconstruction to enhance perceptual naturalness for human listeners in their evaluations, leaving high-frequency details inadequately represented. These shortcomings provide a clear marker for distinguishing synthetic spectrograms from real ones, observable not only to the human eye but also to lightweight machine learning classifiers trained on spectrogram data.

\iffalse
\begin{figure}[h]
  \centering
  \includegraphics[width=\linewidth]{sample-franklin}
  \caption{1907 Franklin Model D roadster. Photograph by Harris \&
    Ewing, Inc. [Public domain], via Wikimedia
    Commons. (\url{https://goo.gl/VLCRBB}).}
  \Description{A woman and a girl in white dresses sit in an open car.}
\end{figure}
\fi

\begin{figure}[h]
  \centering
  \includegraphics[page=8,scale=1]{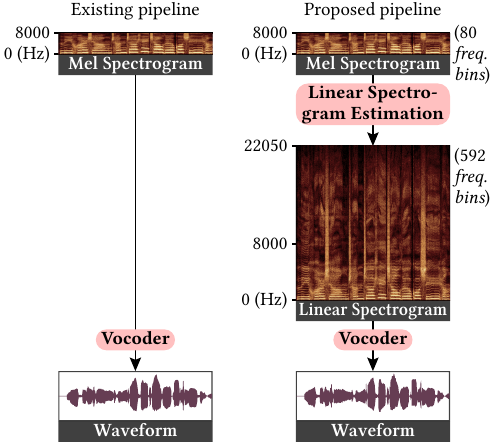}
  \caption{Comparison of the existing vocoder pipeline (left) and the proposed pipeline incorporating linear spectrogram estimation (right) with examples.}
  \Description{Comparison of the existing vocoder pipeline (left) and the proposed pipeline incorporating linear spectrogram estimation (right) with examples.}
  \label{fig:vocoder-pipeline}
\end{figure}

\begin{figure*}[h]
  \centering
  \includegraphics[page=2,scale=1]{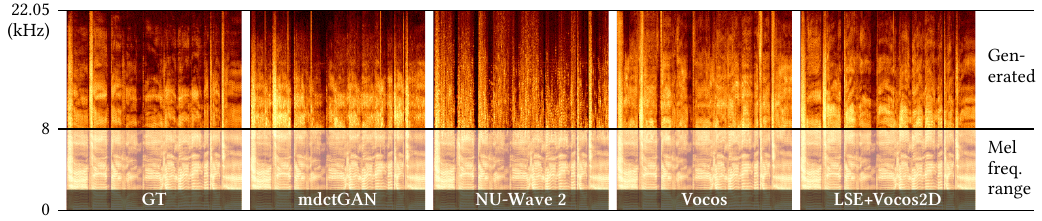}
  \caption{Example of spectrograms generated from existing bandwidth extension methods (mdctGAN \cite{mdctGAN} and NU-Wave 2 \cite{NU-Wave2}) and from vocoder that directly takes Mel spectrogram as input (Vocos \cite{Vocos}), compared with spectrogram generated by our proposed method as well as ground truth spectrogram.}
  \Description{Example of spectrograms generated from existing bandwidth extension methods and from vocoder that directly takes Mel spectrogram as input, compared with spectrogram generated by our proposed method as well as ground truth spectrogram.}
  \label{fig:existing-samples}
\end{figure*}

To address this challenge, we propose an enhancement to the synthesis pipeline by introducing an intermediate step prior to vocoding. Rather than directly feeding Mel spectrograms into the vocoder, we first reconstruct a full-bandwidth linear spectrogram extending up to the Nyquist frequency using a denoising diffusion process \cite{DDPM}. This process utilizes a refined neural network architecture based on DiT \cite{DiT}, meticulously adapted to handle time-frequency spectral data with variable temporal durations. The resulting linear spectrogram is then processed by a reengineered vocoder, termed Vocos2D, which replaces the original 1D convolutions of Vocos \cite{Vocos} with 2D convolutions to effectively manage the expanded frequency bins of linear spectrograms. Figure~\ref{fig:vocoder-pipeline} illustrates the proposed pipeline.

Our approach markedly improves the realism of the generated spectrograms while preserving high audio quality. As demonstrated in Figure~\ref{fig:existing-samples}, the spectrograms generated by our method closely resemble those of real-life recordings, offering a substantial improvement over existing techniques. Objective evaluations, employing lightweight ConvNeXt \cite{ConvNeXt} classifiers trained to detect synthetic spectrograms, demonstrate that our method successfully evades detection. Subjective assessments further reveal that human observers struggle to differentiate our generated spectrograms from real ones. Additionally, Mean Opinion Score (MOS) evaluations indicate a slight enhancement in perceived audio quality with the inclusion of the linear spectrogram estimation step. This work represents a significant leap forward in countering adversarial detection of fake spectrograms generated by vocoders.

\section{Related Work}

\textbf{Bandwidth Extension}, often referred-to as super-resolution, seeks to reconstruct high-frequency components from lower-quality audio inputs, typically transforming waveforms from low to high sampling rates. Notable contributions in this field include AP-BWE \cite{AP-BWE}, UDM+ \cite{UDM+}, mdctGAN \cite{mdctGAN}, and NU-Wave 2 \cite{NU-Wave2}. While these approaches operate directly on waveforms aiming to produce perceptually acceptable audio, they often fail to generate realistic high-frequency spectrograms, leaving them vulnerable to detection when compared to real life recordings, both by trained classifiers and by human eye.

\textbf{Vocoders}. In the domain of vocoder development, especially in terms of generating high-quality audio from Mel spectrograms, methods including PriorGrad \cite{PriorGrad}, DiffWave \cite{DiffWave}, and WaveGrad \cite{WaveGrad} use denoising diffusion processes to iteratively refine audio, while GAN-based vocoders like Vocos \cite{Vocos}, UnivNet \cite{UnivNet}, and HiFi-GAN \cite{HiFi-GAN} leverage adversarial training to produce high-quality audio. However, these models also struggle to accurately reconstruct high-frequency details from Mel spectrogram inputs.

Our linear spectrogram estimation leverages architectural innovations from Transformer-based diffusion models like DiT \cite{DiT} and U-ViT \cite{U-ViT} to better handle time-frequency spectral data with potential long-range dependencies.

Our research also touches on the broader topic of adversarial attacks on fake spectrogram detection. While previously mentioned works focused on generating plausible audio, none have specifically targeted enhancing realism of spectrograms. Our approach maintains audio quality with minimal compromise, while making it more difficult for both human and automated detectors to distinguish between synthetic and real recordings.

\section{Method}

Unlike conventional approaches that directly input a Mel spectrogram into a vocoder, we introduce an intermediate step that estimates a full-bandwidth linear spectrogram from the Mel spectrogram. This linear spectrogram is subsequently processed by a redesigned vocoder, termed Vocos2D, to produce the final audio waveform. Figure~\ref{fig:vocoder-pipeline} juxtaposes the existing synthesis pipeline with our proposed pipeline. 

In view of full-bandwidth linear spectrograms being (1) less common than Mel spectrograms in existing vocoder research while (2) featuring significantly more frequency bins, we cannot expect an existing vocoder to perform well without modifications. To address this, we adapt the high-performing GAN-based Vocos \cite{Vocos} vocoder to create Vocos2D, tailoring it to handle the increased spectral resolution effectively.

The following subsections elaborate on the linear spectrogram specification, the linear spectrogram estimation (LSE) model, and the Vocos2D vocoder.

\subsection{Linear Spectrogram Specification}

Being a pivotal bridge connecting the two processing steps, we first need to define an appropriate specification for the linear spectrogram. On the one hand, we opt against using a log-magnitude spectrogram at full spectral resolution, as it incorporates phase interference from overlapping STFT windows, complicating the task of linear spectrogram estimation, particularly when phase recovery is intended to be vocoder's responsibility. On the other hand, our linear spectrogram should feature equal or finer spectral resolution than the input Mel spectrogram to avoid information loss.  Mel scales are logarithmic and provide finer resolution at lower frequencies. This provides us a maximum distance between adjacent linear filter banks. Moreover, Slaney's implementation \cite{Slaney} of Mel scale is linear below 1000 Hz and logarithmic above 1000 Hz. We could simply extend this linear scale across full frequency range up to the Nyquist frequency. The hop size $\Delta f$ for linear filter banks can then be calculated as:
\begin{equation}
  \Delta f = \mel2hz\mathopen{}\left(\frac{\hz2mel\mathopen{}\left(f_{\text{max}}\right)}{N_{\text{mel}} + 1}\right)\mathclose{},
\end{equation}
where $f_{\text{max}}$ is the cut-off frequency of Mel spectrogram, $N_{\text{mel}}$ is the number of Mel filter banks, and $\mel2hz$ and $\hz2mel$ are functions converting between Mel and Hz scales.

For compatibility with the denoising diffusion process, which requires target data to exhibit zero mean and unit variance, we normalize our spectrogram data. Log magnitude spectrograms typically display a sparse distribution, with values potentially approaching negative infinity upon silence, making them unsuitable as direct denoising targets. To mitigate this, we apply zero mean unit variance normalization with minimum loudness clipping, stabilizing the data for effective denoising.

\subsection{Linear Spectrogram Estimation (LSE) Model Architecture}

% To estimate the linear spectrogram from the input Mel spectrogram, we leverage a denoising diffusion process [1]. Our denoiser model builds on the Diffusion Transformer (DiT) framework [3], with tailored modifications to accommodate the distinct properties of time-frequency data.
We leverage a denoising diffusion process \cite{DDPM} to perform linear spectrogram estimation. Our denoising diffusion model is largely based on the Diffusion Transformer (DiT) \cite{DiT} architecture, with tailored modifications to accommodate the distinct properties of time-frequency data. For instance, images generally maintain a fixed resolution during both training and inference, whereas spectrograms must accommodate variable number of temporal frames while keeping a fixed number of frequency bins. Furthermore, time-frequency data often exhibit varying patterns across frequency spectrum, with low and high frequencies carrying different types of information, rendering spatial invariance applicable along time axis but not frequency axis. To address these challenges, we modified DiT's architecture, focusing on applying mechanisms such as Multi-Head Self-Attention (MHSA) \cite{Transformer} exclusively along time axis, while enabling inter-frequency communication through linear projections with learned embeddings per frequency bin.

\begin{figure}[h]
  \centering
  \includegraphics[page=3,scale=1]{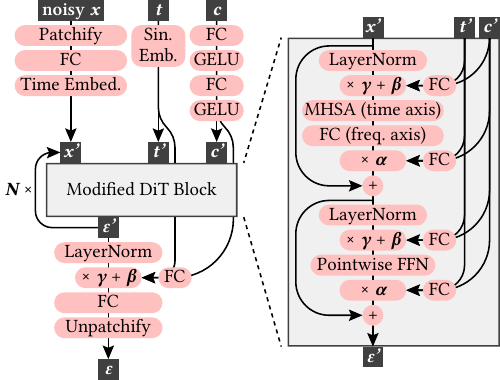}
  \caption{Diagram of the Linear Spectrogram Estimation (LSE) model architecture. FC denotes fully-connected (linear) layer. }
  \Description{Diagram of the Linear Spectrogram Estimation (LSE) model architecture.}
  \label{fig:lse-architecture}
\end{figure}

Figure~\ref{fig:lse-architecture} provides a detailed schematic of our Linear Spectrogram Estimation (LSE) model architecture. The model accepts a noisy spectrogram $x$ with a single channel as its primary input. This noisy $x$ is first downsampled and reorganized into a sequence of patches, akin to the ``patchify'' operation in DiT, effectively converting the 2D spectrogram into a 1D sequence suitable for transformer processing. Then, after a linear projection layer, we apply positional embedding along time axis only, meaning patches form the same time frame share identical embeddings. Transformed patches $x'$ are then processed sequentially through $N$ modified DiT backbone blocks, with the output $\varepsilon'$ of each block serving as the input $x'$ for the next block. At output stage, a corresponding upsampling operation is performed to restore the original spectrogram dimensions, yielding noise residual $\varepsilon$.

\textbf{Conditioning.} Beyond the noisy spectrogram $x$, the LSE model takes two additional inputs: condition $c$ and noise step index $t$. Condition $c$, namely Mel spectrogram, undergoes a series of linear projections with GELU activation along its frequency dimension, producing a transformed condition $c'$ that aligns with the backbone's input requirements. The noise step $t$ is converted into a sinusoidal embedding $t'$, following established practices in diffusion models~\cite{DDPM}, to encode the denoising timestep.

\textbf{Backbone Block.} The backbone block architecture closely follows the adaptive layer normalization block with zero-initialization as used in DiT, featuring two-stage residual processing mechanism. Both stages start with layer normalization, followed by scale-shift operations, and conclude with gate masking. Scale, shift, and gate parameters $\gamma$, $\beta$, and $\alpha$ are regressed from conditions $t'$ and $c'$ using zero-initialized linear layers. The first residual stage employs Multi-Head Self-Attention, applied exclusively along time axis. This allows the model to leverage location invariance in time while treating frames and channels across different frequencies uniformly. Inter-frequency communication is achieved by a subsequent linear projection for accommodating frequency-specific patterns. The second residual stage features a pointwise feed-forward network (FFN), comprising two fully-connected layers with expanded hidden channels, sometimes also referred-to as inverted bottleneck.

\subsection{Vocos2D Design and Training}

The original Vocos \cite{Vocos}, a GAN-based vocoder, is designed to handle Mel spectrograms with a modest number of frequency bins (e.g., 80 or 100), featuring 1D convolutions processing entire frames across all frequencies in its generator. This becomes burdensome when dealing with linear spectrograms with significantly more frequency bins (e.g., over 500). To address this, we draw inspirations from image processing techniques and redesign its generator model to accommodate 2D convolutions, allowing it to focus on specific frequency bands more effectively. Learned embeddings per-frequency is again necessary for inter-frequency communication because location invariance does not hold across frequency bins, as noted previously in LSE model architecture. We also introduce additional shortcut connections from input Mel spectrogram to ConvNeXt blocks so as to enable direct incorporation of the input spectrogram condition.

\begin{figure}[h]
  \centering
  \includegraphics[page=4,scale=1]{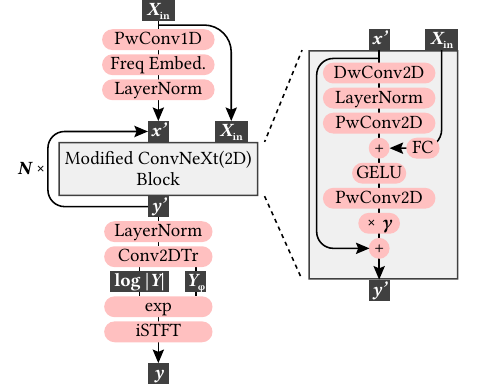}
  \caption{Diagram of Vocos2D generator model architecture. FC denotes fully-connected (linear) layer.}
  \Description{Diagram of Vocos2D generator model architecture.}
  \label{fig:vocos2d-architecture}
\end{figure}

Figure~\ref{fig:vocos2d-architecture} illustrates generator model architecture of Vocos2D vocoder. The input Mel spectrogram condition $X_\text{in}$ first passes through a linear layer, projecting all frequencies of each time frame into a hidden representation, yielding a uniform embedding across frequencies that is subsequently enriched by adding learned frequency embeddings. Following this, layer normalization is applied, producing the intermediate representation $x'$. The transformed $x'$ is then processed sequentially through a series of modified ConvNeXt backbone blocks. The output $y'$ of each block serves as the input $x'$ for the subsequent block. After traversing the backbone, $y'$ undergoes layer normalization, followed by a transposed 2D convolution to upsample the representation to the required number of frequency bins for inverse Short-Time Fourier Transform (iSTFT). This process yields a full-resolution log magnitude spectrogram and a phase spectrogram, which are combined into a complex spectrogram. The final waveform is obtained via iSTFT.

\textbf{Generator Backbone Block.} The backbone block in Vocos2D generator is adapted from the original ConvNeXt block design, augmented with an additional shortcut from $X_{in}$. This shortcut is regressed through a linear layer and added to the inverted bottleneck stage, enhancing the block’s ability to leverage the input condition directly. The core structure of the ConvNeXt block remains intact: depth-wise convolution captures local neighboring information across time and frequency, while point-wise convolution implements the inverted bottleneck with expanded hidden channels. The output is gated by a learned parameter $\gamma$ before adding the residual connection from the input $x'$, producing the block output $y'$. 

\begin{figure}[h]
  \centering
  \includegraphics[page=9,scale=1]{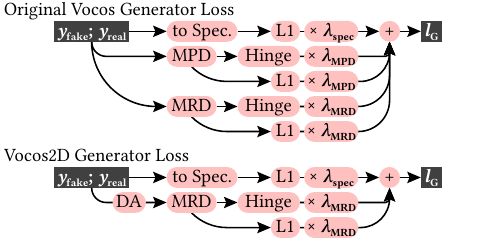}
  \caption{Comparison of the original Vocos generator loss (upper) and the proposed Vocos2D generator loss. DA stands for Discriminator Augmentation \cite{DiffAugment,ADA}. Hinge stands for the Hinge loss \cite{SoundStream}.}
  \Description{Comparison of the original Vocos generator loss (upper) and the proposed Vocos2D generator loss.}
  \label{fig:vocos2d-generator-loss}
\end{figure}

% Vocos2D retains the same discriminator architecture and configuration as the original Vocos.

\textbf{Generator Loss.} A detailed diagram comparing the generator loss for the original Vocos and our proposed Vocos2D is presented in Figure\ref{fig:vocos2d-generator-loss}. We introduce two major modifications to the original Vocos generator loss. First, multi-period discriminator is removed as it exhibited negative impacts on performance when processing input linear spectrograms with increased frequency bins. Second, we incorporate discriminator augmentation (DA) mechanism prior to feeding real and fake waveforms into the multi-resolution discriminator. This augmentation enhances training stability, drawing on prior work \cite{DiffAugment,ADA} that demonstrates improved discriminator convergence with differentiable data transformations applied solely to the discriminator, so that gradients can flow back to the generator through transformations without requiring the generator to actually model undesired augmentations. Specifically, we employ three transformations: random loudness adjustments within $\pm 6$ decibels, random sample shifts, and random phase rotations.

\textbf{Discriminator Training.} Apart from the removal of the multi-period discriminator, the training procedure for the Vocos2D discriminator stays the same as in the original Vocos implementation.

\section{Evaluation}

To thoroughly evaluate the effectiveness of our proposed pipeline, we conducted an extensive series of experiments designed to assess both objective and subjective metrics. Our evaluation framework encompasses both the visual realism of generated spectrograms, as well as the perceived audio quality of synthesized audio.

Initial insights into our approach's performance can be derived from sample spectrograms generated by our proposed LSE+Vocos2D pipeline, presented alongside ground truth spectrograms in Figure~\ref{fig:existing-samples}. Visual inspection reveals that our method achieves markedly superior consistency in high-frequency energy patterns compared to baseline approaches. 

Additional spectrogram samples, as well as listenable audio examples, are accessible at \url{https://lse-vocos2d.github.io/acmmm2025-demo/}. Source code and pretrained models will be made publicly available upon acceptance of this paper.

\subsection{Experimental Setup}

To rigorously evaluate our proposed method, we trained our models on a diverse dataset comprising both privately recorded singing audio and publicly available sources, including OpenCPop~\cite{OpenCPop}, Kiritan~\cite{Kiritan}, CompMusic~\cite{CompMusic}, etc., totaling 106 hours of audio, all resampled to a uniform 44.1 kHz.

\textbf{Spectrogram Specifications.} Input Mel spectrograms were generated using 80 Mel filter banks adhering to Slaney’s implementation~\cite{Slaney}. The linear spectrograms were computed with 592 linear filter banks. Both spectrogram types were calculated at 50 frames per second, equivalent to 0.02 seconds between frames.

\textbf{Linear Spectrogram Estimation (LSE) Model Setup.} The model was configured with 8 backbone blocks, each featuring 8 self-attention heads and a hidden dimension of 320. Training proceeded for 1,200,000 steps with a batch size of 18, utilizing AdamW~\cite{AdamW} as optimizer at an initial learning rate of $10^{-4}$ under FP16 precision. All training audio clips were segmented into 8-second durations. Learning rate halved whenever the loss failed to reach a new minimum over 150,000 consecutive steps. During inference, we employed DPM++ 2M Karras~\cite{DPM-Solver} as noise sampler with 32 sampling steps.

\textbf{Vocos2D and Baseline Vocos Configurations.} The Vocos2D generator model, tailored for linear spectrograms, was equipped with 24 backbone blocks and a hidden dimension of 256. The baseline Vocos generator model~\cite{Vocos}, designed for Mel spectrograms, was configured with 10 backbone blocks and a hidden dimension of 512. Both models were trained for 900,000 steps using AdamW as optimizer at an initial learning rate of $5 \times 10^{-4}$. Learning rate decayed exponentially at a rate of $0.995$. Training audio for both models was segmented into 4-second clips. Additionally, saved checkpoints for both models utilized exponential moving average (EMA) with a decay rate of $0.999$ to promote training stability.

\subsection{Objective Visual Evaluation on Spectrograms}

Our evaluation begins with an objective analysis on visual realism of generated spectrograms, employing a lightweight ConvNeXt~\cite{ConvNeXt} model as a visual discriminator to distinguish between real and synthetic spectrograms. The model architecture comprises 8 ConvNeXt blocks with downsampling ratios of [4, 4, 2, 2]. We randomly sampled 18 hours of singing voice recordings from the aforementioned data corpus, allocating 16 hours for training and 2 hours for testing. This setup enabled us to assess how effectively our generated spectrograms could deceive the discriminator into classifying them as real.

\begin{figure}[h]
  \centering
  \includegraphics[page=5,scale=1]{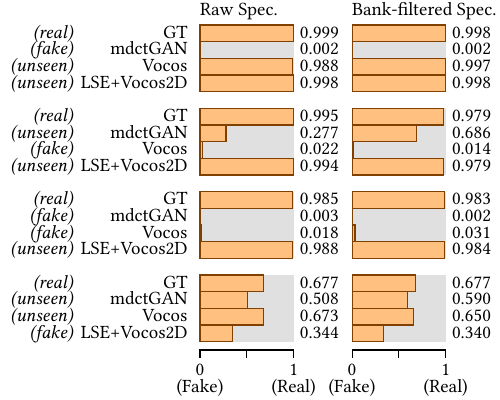}
  \caption{Classification results from all 8 trained discriminator models. All results were run on test samples disjoint from training data. Rows marked with ``real'' or ``fake'' imply seen during training.}
  \Description{Classification results from all 8 trained discriminator models.}
  \label{fig:eval-visual-objective}
\end{figure}

We trained 8 instances of the discriminator under varying conditions to explore different configurations of negative training samples. The classification results are presented in Figure~\ref{fig:eval-visual-objective}, with row prefixes indicating whether samples were seen during training (labeled ``real'' or ``fake'') or unseen (labeled ``unseen''). We evaluated two spectrogram input types: raw log magnitude spectrograms and spectrograms filtered by linear filter banks. Both types yielded similar outcomes. All discriminator instances used real linear spectrograms as positive ground truth (GT). Four distinct settings for negative training samples were investigated:
\begin{itemize}
\item \textbf{mdctGAN only}: Both Vocos and LSE+Vocos2D (proposed) were classified as nearly real, with LSE+Vocos2D achieving marginally higher.
\item \textbf{Vocos only}: LSE+Vocos2D remained nearly indistinguishable from GT, while mdctGAN scored significantly lower.
\item \textbf{Both mdctGAN and Vocos}: LSE+Vocos2D consistently scored close to GT, outperforming both baselines.
\item \textbf{LSE+Vocos2D only}: The discriminator struggled to differentiate LSE+Vocos2D from GT, with all scores hovering around 0.5, suggesting near-random guessing.
\end{itemize}

These results indicate that spectrograms generated by our proposed LSE+Vocos2D pipeline are significantly more challenging to distinguish from real spectrograms compared to baseline approaches. However, we observed a slight bias in the discriminator toward classifying unseen samples as real, possibly due to the limited diversity of baseline methods included into the training data. To provide a more holistic assessment, we complement these findings with subjective evaluations in the subsequent subsections.

\subsection{Subjective Visual Evaluation on Spectrograms}

To further evaluate spectrogram realism from a human perspective, we conducted a subjective study involving 216 spectrogram images generated from 36 randomly selected audio samples. These included spectrograms produced by mdctGAN, NU-Wave 2, Vocos, our proposed LSE+Vocos2D pipeline, and ground truth (GT) audio. Participants, all aged between 20 and 35 with at least amateur experience in vocal audio processing, were asked to identify whether each spectrogram originated from a real recording or synthetic audio. Responses from participants who rated GT spectrograms lower than others were discarded to ensure response reliability.

\begin{figure}[h]
  \centering
  \includegraphics[page=6,scale=1]{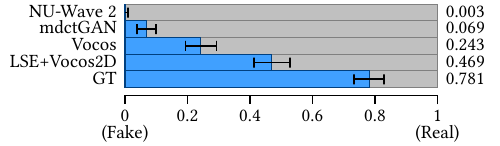}
  \caption{Human evaluation statistics for visual realism of spectrograms. Error bars represent 95\% confidence interval.}
  \Description{Human evaluation statistics for visual realism of spectrograms.}
  \label{fig:eval-visual-subjective}
\end{figure}

% Valid responses were collected from 8 participants. 
Result statistics are illustrated in Figure~\ref{fig:eval-visual-subjective}. Error bars represent 95\% confidence interval. Spectrograms generated by LSE+Vocos2D were perceived as real 46.9\% of the time, the highest among synthetic methods, below GT scoring 78.1\%. In comparison, Vocos spectrograms were identified as real only 24.3\% of the time, while NU-Wave 2 and mdctGAN scored both below 10\%. These results highlight the substantial improvement in visual realism achieved by our proposed method, narrowing the perceptual gap between synthetic and authentic spectrograms.

\subsection{Perceived Audio Quality Evaluation}

To ensure that enhancements in spectrogram realism do not compromise audio quality, we conducted a comprehensive evaluation of perceived audio quality using both objective and subjective methods. This assessment involved 180 audio clips derived from the same 36 audio pieces used in the previous subjective spectrogram evaluation. The evaluation framework incorporated three objective systems:
\begin{itemize}
\item UTMOS~\cite{UTMOS}: An automatic Mean Opinion Score (MOS) prediction system for speech and audio quality assessment.
\item NISQA~\cite{NISQA}: Also an automatic speech quality prediction model estimating MOS score.
\item Meta Audiobox Aesthetics \cite{AudioboxAesthetics}: A unified automatic quality assessment tool for speech, music, and sound, from which we used the production quality (PQ) metric.
\end{itemize}

We also conducted a subjective MOS evaluation, where participants rated audio quality on a scale from 1 to 5. Responses were filtered to exclude those who rated unprocessed GT audio below 5 or synthetic audio above 4 more than 50\% of the time, ensuring response reliability. 
%Participants included those from the previous subjective spectrogram evaluation and additional individuals from online virtual singer communities, all aged between 20 and 35. Valid responses were collected from 15 participants.

\begin{figure}[h]
  \centering
  \includegraphics[page=1,scale=1]{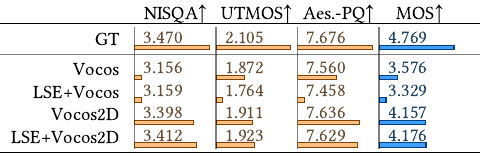}
  \caption{Audio quality evaluation results. Metrics colored in orange denote objective evaluation; metrics colored in blue denote subjective evaluation.}
  \Description{Audio quality evaluation results.}
  \label{fig:eval-audio}
\end{figure}

Evaluation results are summarized in Figure~\ref{fig:eval-audio}. Across all objective metrics, GT audio scored highest, confirming the reliability of the evaluation systems. Key findings include:
\begin{itemize}
\item \textbf{Baseline Vocos with LSE}: Enabling LSE slightly degraded audio quality compared to Vocos without LSE. This is likely because Vocos sees the input spectrogram only once at the beginning, and relies on memorizing spectrogram information across layers, a process strained by the increased frequency bins in reconstructed linear spectrograms.
\item \textbf{Vocos2D with LSE}: In contrast, Vocos2D with LSE achieved marginally higher scores overall compared to plain Vocos2D, benefiting from shortcut connections that propagate input spectrogram information directly to backbone blocks, effectively leveraging the additional spectral detail.
\end{itemize}

Note that both Vocos and Vocos2D showed slight improvements with LSE enabled under NISQA, likely because NISQA takes into account frequencies up to 20 kHz, whereas UTMOS and Audiobox Aesthetics resample inputs to 16 kHz, limiting their spectral scope to 8 kHz.

These findings confirm that our proposed LSE+Vocos2D pipeline could maintain a high standard of perceived output audio quality despite the additional processing step.

\section{Conclusion}

In this paper, we present a novel approach to elevate the realism of vocoder-generated singing voice audio by addressing the persistent disparities between synthetic and real spectrograms, particularly in high-frequency regions. Our proposed pipeline integrates two key innovations: an explicit Linear Spectrogram Estimation (LSE) step, leveraging a denoising diffusion process with a tailored DiT-based architecture, and Vocos2D, a redesigned vocoder optimized for linear spectrograms with increased frequency resolution. Through rigorous objective and subjective evaluations, we demonstrate that LSE+Vocos2D produces spectrograms that consistently deceive both lightweight neural discriminators and human observers, achieving classification scores approaching those of ground truth recordings. Notably, this enhanced realism is achieved with minimal impact on perceived audio quality, as evidenced by subjective MOS scores and objective metrics that remain competitive with baseline methods. These findings underscore the efficacy of our method in overcoming longstanding limitations in vocoding techniques, particularly in the context of adversarial detection of synthetic spectrograms.

\section*{Declaration of Generative AI and AI-assisted Technologies in the Writing Process}
During the preparation of this work, the authors utilized a combination of ChatGPT 4o, DeepSeek R1 and Grok 3 solely for the purpose of linguistic enhancement and grammar checking. After using these tools/services, the authors thoroughly reviewed and edited the content as needed and take full responsibility for the content of the publication.

% \begin{acks}
% \end{acks}

\bibliographystyle{ACM-Reference-Format}
\bibliography{paper-lse}

\iffalse
\appendix

\section{Research Methods}

\subsection{Part One}

Lorem ipsum dolor sit amet, consectetur adipiscing elit. Morbi
malesuada, quam in pulvinar varius, metus nunc fermentum urna, id
sollicitudin purus odio sit amet enim. Aliquam ullamcorper eu ipsum
vel mollis. Curabitur quis dictum nisl. Phasellus vel semper risus, et
lacinia dolor. Integer ultricies commodo sem nec semper.

\subsection{Part Two}

Etiam commodo feugiat nisl pulvinar pellentesque. Etiam auctor sodales
ligula, non varius nibh pulvinar semper. Suspendisse nec lectus non
ipsum convallis congue hendrerit vitae sapien. Donec at laoreet
eros. Vivamus non purus placerat, scelerisque diam eu, cursus
ante. Etiam aliquam tortor auctor efficitur mattis.

\section{Online Resources}

Nam id fermentum dui. Suspendisse sagittis tortor a nulla mollis, in
pulvinar ex pretium. Sed interdum orci quis metus euismod, et sagittis
enim maximus. Vestibulum gravida massa ut felis suscipit
congue. Quisque mattis elit a risus ultrices commodo venenatis eget
dui. Etiam sagittis eleifend elementum.

Nam interdum magna at lectus dignissim, ac dignissim lorem
rhoncus. Maecenas eu arcu ac neque placerat aliquam. Nunc pulvinar
massa et mattis lacinia.
\fi

\end{document}